\documentclass[12pt]{article}
\usepackage{graphicx}

\newcommand{\beq}{\begin{equation}}
\newcommand{\eeq}{\end{equation}}
\newcommand{\beqn}{\begin{eqnarray}}
\newcommand{\eeqn}{\end{eqnarray}}
\newcommand{\beqns}{\begin{eqnarray*}}
\newcommand{\eeqns}{\end{eqnarray*}}

\begin{document}
\begin{titlepage}
\begin{center}

\hfill USTC-ICTS-06-13\\
\hfill September 2006\\

\vspace{2.5cm}

{\Large {\bf   Strong phases, asymmetries, and SU(3) symmetry
breaking in $D\to
 K \pi$  decays}} \vspace*{1.0cm}\\
{  Dao-Neng Gao$^\dagger$} \vspace*{0.3cm} \\
{\it\small Arnold Sommerfeld Center, Department f\"ur Physik,
Ludwig-Maximilians-Universit\"at M\"unchen,  Theresienstr. 37,
D-80333, Munich, Germany} \\{\it\small and} \\{\it\small
Interdisciplinary Center for Theoretical Study and Department of
Modern Physics, University of Science and Technology of China,
Hefei, Anhui 230026 China}\vspace*{1cm}
\end{center}
\begin{abstract}
\noindent Motivated by some new experimental data, we carry out a
phenomenological analysis of $D\to K\pi$ decays including both
Cabibbo favored and doubly Cabibbo suppressed modes. Two
asymmetries, $R(D^0)$ and $R(D^+)$, which are generated through
interference between Cabbibo favored and doubly Cabibbo suppressed
$D\to K\pi$ transitions, are predicted. The relative strong phase,
$\delta_{K\pi}$, between $D^0\to K^-\pi^+$ and $D^0\to K^+\pi^-$
decays, is estimated. The theoretical results agree well with the
current measurements.

\end{abstract}

\vfill \noindent

$^{\dagger}$ Email address: ~gaodn@ustc.edu.cn
\end{titlepage}

\section{Introduction}

Non-leptonic $D\to K\pi$ decays and their strong phases have been
of great interest as they are essentially related to the studies
of CP violation, $D^0-\bar{D}^0$ mixing, and SU(3) symmetry
breaking effects in charm physics \cite{FNP99, GGR01, CR9902,
CR02, CLR03}.  These decay modes contain both Cabibbo favored (CF)
and doubly Cabibbo suppressed (DCS) transitions, and the effective
Hamiltonian relevant for them is given by \beqn\label{Hamiltonian}
{\cal H}_{\rm eff}&=&\frac{G_F}{\sqrt{2}}\left\{V_{ud}V^*_{cs}
[C_1(\bar{s}_i c_i)_{V-A}(\bar{u}_j d_j)_{V-A}+C_2 (\bar{s}_i
c_j)_{V-A}(\bar{u}_j
d_i)_{V-A}]\right.\nonumber\\&&\left.+V_{us}V^*_{cd}[C_1(\bar{d}_i
c_i)_{V-A}(\bar{u}_j s_j)_{V-A}+C_2 (\bar{d}_i
c_j)_{V-A}(\bar{u}_j s_i)_{V-A}]\right\}\nonumber\\&&+{\rm H.c.},
\eeqn where $V-A$ denotes $\gamma_\mu(1-\gamma_5)$. The first line
in eq. (\ref{Hamiltonian}) governs CF decays and the second line
DCS decays.

Theoretically, factorization hypothesis has been widely utilized
in the hadronic $D$ decays. Many studies are based on the naive
factorization approach, which simply replaces the matrix elements
of a four-fermion operator in a heavy-quark decay by the product
of the matrix elements of two currents. This approach has long
been used in phenomenological applications, although there is an
obvious shortcoming that it cannot lead to the scale and scheme
independence for the final physical amplitude. Several years ago,
the authors of Ref. \cite{BBNS} have formed an interesting QCD
factorization formula for the two-body exclusive non-leptonic $B$
decays, in which the scale and scheme dependence of the hadronic
matrix elements is recovered, and the naive factorization can be
obtained as the lowest order approximation. The radiative
corrections in the strong coupling constant $\alpha_s$, which are
dominated by hard gluon exchange, can be calculated systematically
using the perturbative QCD in the heavy quark limit. This means
the strong final-state interaction phases, which arise from the
hard-scattering kernel, are calculable from first principles.
Analogously, in the heavy charm quark limit, a similar
factorization formula for the matrix elements of the operators
$Q_i$'s in the effective weak Hamiltonian (\ref{Hamiltonian}) can
be  written as \cite{BBNS} \beqn\label{FF}\langle P_1
P_2|Q_i|D\rangle =\sum_j F_j^{D\to P_1}(m_2^2)\int^1_0 du
T^I_{ij}(u)\phi_{P_2}(u)+(P_1\leftrightarrow P_2)\nonumber\\
+\int^1_0 d\xi du dv
T_i^{II}(\xi,u,v)\phi_D(\xi)\phi_{P_1}(v)\phi_{P_2}(u), \eeqn
where $F_j^{D\to P_{1,2}}(m_{1,2}^2)$ denotes a $D\to P_{1,2}$
form factor, $\phi_X(u)$ is the light-cone distribution amplitude
of meson $X$. $T^I_{ij}(\xi,u,v)$ and $T^{II}_i(\xi,u,v)$ are
hard-scattering functions, which are perturbatively calculable.
Then theoretical results for $D$ decays can be obtained
straightforwardly. Taking the CF decays $D^0\to K^-\pi^+$, $D^0\to
\bar{K}^0 \pi^0$, and $D^+\to \bar{K}^0\pi^+$ as examples, for the
leading power contribution, we get \beq{\cal B}(D^0\to
K^-\pi^+)=3.97\%,\;\; {\cal B}(D^0\to \bar{K}^0 \pi^0)=0.08\%,\;\;
{\cal B}(D^+\to \bar{K}^0\pi^+)=7.66\% \eeq at the scale $\mu=1.5$
GeV. (Here we have parameterized $\int^1_0 d\xi~ \phi_D(\xi)/\xi
\equiv m_D/\lambda_D$ and set $\lambda_D=0.3$ GeV in the numerical
calculations.) The corresponding experimental data from
\cite{PDG06} are
$$ {\cal B}(D^0\to K^-\pi^+)=(3.80\pm 0.07)\%,\;\; {\cal B}(D^0\to
\bar{K}^0 \pi^0)=(2.28\pm 0.24)\%,$$
$${\cal B}(D^+\to \bar{K}^0\pi^+)=(2.94\pm 0.12)\%.$$  It is seen
that, although the predicted branching ratio for the color-allowed
decay $D^0\to K^-\pi^+$ is in qualitative agreement with the data,
the prediction for the color-suppressed decay $D^0\to \bar{K}^0
\pi^0$ is too small, and for the charged mode, the theoretical
${\cal B}(D^+\to \bar{K}^0\pi^+)$ is too large. Similar conclusion
will be reached when applying the formula (\ref{FF}) to singly
Cabibbo-suppressed (SCS) and DCS decays. This seems to indicate
that the charm quark mass is not heavy enough to apply the QCD
factorization approach \cite{BBNS} or pQCD \cite{KLS} in $D$
decays very reliably. Therefore one generally appeals to the
phenomenological analysis of these processes.

Experimentally, many new results in $D$ decays are expected soon
from the dedicated experiments conducted at CLEO, E791, FOCUS,
SELEX, and the $B$ factories BaBar and Belle.  In particular, as
pointed out in \cite{BY95, BLMPS}, there are interesting
asymmetries due to interference between CF and DCS $D\to K\pi$
transitions, defined as  \beq\label{asy0} R(D)\equiv \frac{{\cal
B}(D\to K_S \pi)-{\cal B}(D\to K_L \pi)}{{\cal B}(D\to K_S
\pi)+{\cal B}(D\to K_L \pi)},\eeq which have been observed by CLEO
Collaboration \cite{CLEO06} very recently,
\beq\label{asy1}R(D^0)=0.122\pm 0.024\pm 0.030, \;\;\;
R(D^+)=0.030\pm 0.023\pm0.025. \eeq  Also a preliminary result on
the relative strong phase between $D^0\to K^-\pi^+$ and $D^0\to
K^+\pi^-$, which is due to SU(3) symmetry breaking and important
in the search for $D^0-$ $\bar{D}^0$ mixing \cite{FNP99, Belle06},
has been reported by CLEO Collaboration as \beq\label{deltakpiexp}
\cos\delta_{ K\pi}=1.09\pm0.66~ \cite{Sun06},\eeq although with
very large uncertainty.

Motivated by the new measurements mentioned above, we would like
to perform a phenomenological analysis of both CF and DCS $D\to
K\pi$ decays. As will be shown below, the present data cannot
allow us to determine all of the phenomenological parameters
appearing in decay amplitudes. Implementing the SU(3) symmetry may
constrain the amplitudes, thus largely reduce the number of
independent parameters. However, it is known that this symmetry is
not well respected in nature, even badly broken in some cases.
Therefore, as a conservative way to constrain these amplitudes, in
this paper we assume that SU(3) symmetry in $D\to K\pi$ decays is
{\it moderately} broken, namely, symmetry breaking effects in
decay amplitudes are dominated by decay constants $f_P$ and $D\to
P$ ($P=\pi,~K$) form factors, and other SU(3) symmetry breaking
sources can be neglected. This is not a general feature of SU(3)
symmetry breaking in charmed decays.

\section{$D\to K\pi$ decay amplitudes}

We begin by considering the $D\to K\pi$ decay amplitudes in terms
of the quark-diagram topologies ${\cal T}$ (color-allowed), ${\cal
C}$ (color-suppressed), ${\cal E}$ ($W$-exchange), and ${\cal A}$
($W$-annihilation) \cite{CC86}, which are given by
\beqn\label{topoamplitude1} A(D^0\to
K^-\pi^+)=i\frac{G_F}{\sqrt{2}}V_{ud}V^*_{cs}({\cal T}+{\cal
E}),\\
\sqrt{2} A(D^0\to \bar{K}^0
\pi^0)=i\frac{G_F}{\sqrt{2}}V_{ud}V^*_{cs}({\cal
C}-{\cal E}),\\
A(D^+\to \bar{K}^0\pi^+)=i\frac{G_F}{\sqrt{2}}V_{ud}V^*_{cs}({\cal
T}+{\cal C}),\\
A(D^0\to K^+\pi^-)=i\frac{G_F}{\sqrt{2}}V_{us}V_{cd}^*({\cal
T}^\prime+{\cal E}^\prime),\\
\sqrt{2}A(D^0\to K^0
\pi^0)=i\frac{G_F}{\sqrt{2}}V_{us}V_{cd}^*({\cal C}^\prime-{\cal
E}^{\prime}),\\
A(D^+\to K^0\pi^+)=i\frac{G_F}{\sqrt{2}}V_{us}V_{cd}^*({\cal
C}^{\prime}+{\cal A}^\prime),\\
\label{topoamplitude7}\sqrt{2}A(D^+\to
K^+\pi^0)=i\frac{G_F}{\sqrt{2}}V_{us}V_{cd}^*({\cal
T}^\prime-{\cal A}^\prime),\eeqn and two isospin relations
\beq\label{isocf} A(D^0\to K^- \pi^+)+\sqrt{2} A(D^0\to \bar{K}^0
\pi^0=A(D^+\to \bar{K}^0\pi^+),\eeq  \beq\label{isodcs} A(D^0\to
K^+\pi^-)+\sqrt{2} A(D^0\to K^0\pi^0)=A(D^+\to K^0 \pi^+)+\sqrt{2}
A(D^+\to K^+\pi^0)\eeq are satisfied explicitly. For our
notations, we have extracted the CKM matrix elements and factor
$G_F/\sqrt{2}$ from the quark-diagram amplitudes, and the prime is
added to DCS amplitudes.

The present experimental status of $D\to K\pi$ decays is not very
satisfying. Branching ratios of three CF modes and ${\cal
B}(D^0\to K^+\pi^-)$ have been reported by Particle data group
\cite{PDG06}, however, there are no measurements for ${\cal
B}(D^0\to K^0\pi^0)$ and ${\cal B}(D^+\to K^0 \pi^+)$ yet. Only an
upper bound on ${\cal B}(D^+\to K^+\pi^0)<4.2\times 10^{-4}$
(CL=90\%) is shown in \cite{PDG06}; while, very recently, BaBar
Collaboration and CLEO Collaboration have given \beqn\label{Babar}
{\cal B}(D^+\to K^+\pi^0)=(2.52\pm
0.47\pm 0.25\pm 0.08)\times 10^{-4} ~~\cite{BABAR06},\nonumber\\
\\\nonumber
{\cal B}(D^+\to K^+\pi^0)=(2.25\pm 0.36\pm0.15\pm0.07)\times
10^{-4} ~~\cite{CLEO0607},\eeqn respectively.  In general the
quark-diagram amplitudes in (\ref{topoamplitude1}) --
(\ref{topoamplitude7}) could have non-trivial strong phases.
Therefore,  only using the available experimental data, it is
impossible to determine these amplitudes without any theoretical
assumptions.

On the other hand, with the help of the factorization hypothesis,
${\cal T}$, ${\cal T}^\prime$, ${\cal C}$, and ${\cal C}^\prime$
can be expressed
 as \beqn\label{factorization1}{\cal T}=f_\pi(m_D^2-m_K^2) F^{D\to
 K}_0(m_\pi^2) a_1^{\rm eff},\nonumber\\ {\cal C}=f_K(m_D^2-m_\pi^2)
 F^{D\to\pi}_0(m_K^2) a_2^{\rm eff},\nonumber\\
{\cal T}^\prime=f_K(m_D^2-m_\pi^2) F^{D\to \pi}_0(m_K^2) a_1^{\rm
eff},\nonumber\\{\cal C}^\prime=f_K(m_D^2-m_\pi^2)
 F^{D\to\pi}_0(m_K^2) a_2^{\rm eff},
 \eeqn
where $a_i^{\rm eff}$'s are  regarded as the effective Wilson
coefficients fixed from the data (in the naive factorization,
$a_{1,2}=C_{1,2}+C_{2,1}/N_c$). Generally in the factorization
approach $a_i^{\rm eff}$'s in DCS amplitudes could be different
from the ones in CF amplitudes due to SU(3) symmetry breaking
effects. Here we do not distinguish them because, as stated above,
we assume that the SU(3) symmetry breaking effects in $D\to K \pi$
modes have been mostly captured by the decays constants ($f_\pi$
and $f_K$) and form factors ($F^{D\to \pi, K}_0(q^2)$) in the
amplitudes.

\begin{figure}[t]
\begin{center}
\includegraphics[width=14cm,height=3.0cm]{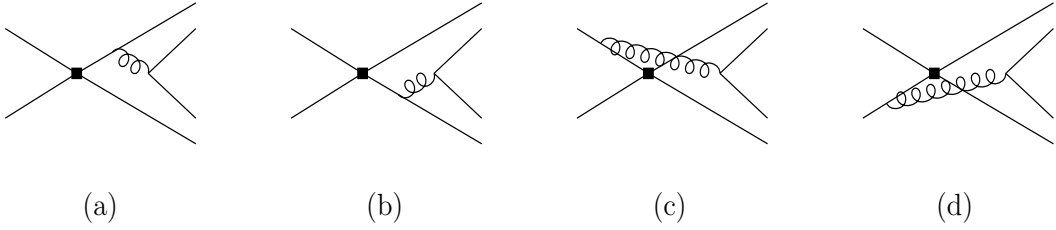}
\end{center}
\caption{Weak annihilation diagrams ($W$-exchange or
$W$-annihilation) via gluon emission. The solid square denotes the
weak vertex.}
\end{figure}

A similar analysis of $W$-exchange and $W$-annihilation amplitudes
leads to \beqn\label{factorization2}{\cal E}=f_D(m_K^2-m_\pi^2)
F_0^{0\to
K\pi}(m_D^2) a_2^{\rm eff},\nonumber\\
{\cal E}^\prime=f_D(m_\pi^2-m_K^2) F_0^{0\to K\pi}(m_D^2)
a_2^{\rm eff},\nonumber\\
{\cal A}^\prime=f_D(m_\pi^2-m_K^2) F_0^{0\to K\pi}(m_D^2) a_1^{\rm
eff}.\eeqn This will give ${\cal E}=-{\cal E}^\prime$. However,
the annihilation form factor $F^{0\to K\pi}_0(m_D^2)$ is expected
to be strongly suppressed due to the large $q^2=m_D^2$
\cite{LB79}, therefore contributions from eq.
(\ref{factorization2}) are believed to be negligible. For $B$
mesons, the weak annihilation amplitudes ($W$-exchange or
$W$-annihilation) induced by the topologies of gluon emission
arising from the quarks of the weak vertex, as shown in Fig. 1,
have been analyzed in \cite{BBNS01}, which is thought to be
numerically important in $B$ decays. The similar study on $D$
mesons have been done in \cite{CZ82,LY05}, and it has been shown
that these contributions could also play important roles in $D$
decays. The $O(\alpha_s)$ contribution can be read directly from
Refs. \cite{BBNS01, LY05}, \beq\label{annihilation1} {\cal E}=f_D
f_K f_\pi\frac{C_F}{N_C^2}\pi \alpha_s
C_1\left[18\left(X_A-4+\frac{\pi^2}{3}\right)+2 r_\chi^\pi
r_\chi^K X_A^2\right], \;\;\;\;\; {\cal E}^\prime={\cal E},\eeq
and \beq\label{annihilation2} {\cal A}^\prime=f_D f_K
f_\pi\frac{C_F}{N_C^2}\pi \alpha_s
C_2\left[18\left(X_A-4+\frac{\pi^2}{3}\right)+2 r_\chi^\pi
r_\chi^K X_A^2\right].\eeq
 where $X_A=\int^1_0 dy/y$ has been
used to parameterize the logarithmically  divergent integrals due
to the end-point singularity, and $C_1$, $C_2$ are the Wilson
coefficients in (\ref{Hamiltonian}). Note that the asymptotic form
of the light-cone distribution amplitudes for light mesons have
been used in the derivation of eq. (\ref{annihilation1}). This is
consistent with the assumption on SU(3) symmetry breaking used in
deriving eq. (\ref{factorization1}).

Meanwhile, comparing eq. (\ref{annihilation2})  with  eq.
(\ref{annihilation1}), one can get an additional constraint
\beq\label{annihilation3} {\cal A}^\prime=\frac{C_2}{C_1}{\cal
E}^\prime.\eeq We would like to give some remarks here.

\begin{itemize}

\item By combining eq. (\ref{annihilation3}) with eqs.
(\ref{annihilation1}) and (\ref{factorization1}), we will reduce
independent complex phenomenological parameters appearing in decay
amplitudes as $a_1^{\rm eff}$, $a_2^{\rm eff}$, and ${\cal E}$.
Since only five branching ratios of $D\to K\pi$ decays are
measured up to now, this means that including the additional
constraint (\ref{annihilation3}) is important to enable us to
determine the $D\to K\pi$ amplitudes from the present data. From
\cite{BBL}, $C_1$ and $C_2$  have opposite signs, therefore
weak-exchange ${\cal E}$ and weak-annihilation ${\cal A}$
amplitudes have opposite signs, consistent with the observations
in \cite{cheng03}. Also $|C_2|<|C_1|$ \cite{BBL}, we have $|{\cal
E}^\prime|>|{\cal A}^\prime|$; while the contrary conclusion will
be obtained if we use eq. (\ref{factorization2}).

\item Strictly speaking, we have to admit that eq.
(\ref{annihilation3}) is not very physical since $C_1$ and $C_2$
are both scale and scheme dependent \cite{BBL}. The scale and
scheme dependence of $C_2/C_1$ has been shown in Table 1, from
which it is found that $C_2/C_1$ is about $-0.5\sim -0.3$ for
$\mu$ around $1.0 \sim 1.5$ GeV (Note that the scale in this range
is relevant for $D$ decays). Therefore, we will treat in the
following numerical calculations $C_2/C_1$ as a negative parameter
instead of a ratio of two Wilson coefficients.

\item The weak-annihilation contribution is power suppressed in
the heavy quark limit. The divergent integral $X_A$ appearing in
(\ref{annihilation1}) and (\ref{annihilation2}) signals that
factorization breaks down actually. In the present analysis we
therefore use relations ${\cal E}={\cal E}^\prime$ and
(\ref{annihilation3}) for these three weak-annihilation amplitudes
instead of their explicit expressions shown above. Although they
are not model independent relations, one will find that
phenomenologically they work very well in $D\to K\pi$ decays.

\end{itemize}

\begin{table}[t]\begin{center}\begin{tabular}{c| c c c}\hline\hline &&&\\
 & $\Lambda^{(4)}_{\overline{\rm MS}}=215 ~{\rm
MeV}$&$\Lambda^{(4)}_{\overline{\rm MS}}=325 ~{\rm MeV}$&
$\Lambda^{(4)}_{\overline{\rm MS}}=435 ~{\rm MeV}$\\\\\hline\\
$\mu$ [GeV]& NDR ~~~~  HV & NDR ~~~~ HV & NDR ~~~~ HV
\\\\ 1.0 & -0.339 ~~ -0.390 & -0.400 ~~ -0.464 & -0.464 ~~ -0.548
\\1.5& -0.277 ~~ -0.319 & -0.318 ~~ -0.369& -0.357 ~~ -0.419 \\
2.0 & -0.240 ~~ -0.278 & -0.271 ~~ -0.316 & -0.301 ~~ -0.353\\
\hline\hline
\end{tabular}\caption{The scale and scheme dependence of $C_2/C_1$ at the next-to-leading order.
The values of $C_1$ and $C_2$ are taken directly from
\cite{BBL}.}\end{center}\end{table}

\section{Phenomenological analysis}

From now on we study some possible phenomenological applications
based on the above theoretical assumptions.

First, the use of eqs. (\ref{factorization1}) and
(\ref{annihilation1}) gives ${\cal C}={\cal C}^\prime$ and ${\cal
E}={\cal E}^\prime$, hence we have \beq \label{16}{A(D^0\to {K}^0
\pi^0)}=\frac{V_{us}V_{cd}^*}{V_{ud} V_{cs}^*} A(D^0\to \bar{K}^0
\pi^0)=-\tan^2\theta_C A(D^0\to \bar{K}^0 \pi^0),\eeq which
implies that the relative strong phase between these two
amplitudes vanishes. Here $\theta_C$ is the Cabibbo angle.
Consequently, one gets \beq \label{rd0} R(D^0)=\frac{2
\tan^2\theta_C}{1+\tan^4 \theta_C}\simeq 2 \tan^2\theta_C. \eeq
Using $\tan\theta_C\simeq 0.23$, $R(D^0)\simeq 0.106$, which is in
agreement with the measurement in eq. (\ref{asy1}). The same
result has been obtained in Refs. \cite{BY95, JR06}. However, one
cannot expect the similar result as (\ref{rd0}) for $R(D^+)$ since
there is no similar relation as (\ref{16}) between $A(D^+ \to
\bar{K}^0 \pi^+)$ and $A(D^+\to K^0 \pi^+)$, even in the SU(3)
symmetry limit. We will discuss this issue later.

Second, using the constraint (\ref{annihilation3}) together with
(\ref{annihilation2}) and (\ref{factorization1}), one can obtain
an interesting relation among the amplitudes of $D^0\to K^\pm
\pi^\mp$ and $D^+\to K^+\pi^0$ decays, which is
\beq\label{sumrules}\tan^2\theta_C A(D^0\to K^-\pi^+)+ \kappa
A(D^0\to K^+\pi^-)=\sqrt{2}\zeta A(D^+\to K^+\pi^0),\eeq where
\beqn\label{kappa} \kappa\equiv\frac{1+C_{2}/C_1~
x~f_{\pi}/f_{K}}{1+C_{2}/C_1},\;\;\;\zeta\equiv\frac{1-x~
f_{\pi}/f_ K}{1+C_{2}/C_1}\nonumber \eeqn  with \beqn\label{xx}
x\equiv\frac{(m_D^2-m_K^2) F_0^{D\to
K}(m_\pi^2)}{(m_D^2-m_\pi^2)F_0^{D\to \pi}(m_K^2)}. \eeqn    Thus
the relative strong phase $\delta_{K\pi}$ between $A(D^0\to
K^+\pi^-)$ and $A(D^0\to K^-\pi^+)$ is given by
\beqn\label{deltakpi} \cos\delta_{K\pi}=\frac{\tan^4\theta_C {\cal
B}(D^0\to K^-\pi^+)+ \kappa^2 {\cal B}(D^0\to K^+\pi^-)-2
\zeta^2\frac{\tau(D^0)}{\tau(D^+)}{\cal B}(D^+\to K^+\pi^0)}{2
\tan^2\theta_C \kappa\sqrt{{\cal B}(D^0\to K^-\pi^+) {\cal
B}(D^0\to K^+\pi^-)}},\eeqn  where $\tau(D)$ is the life time of
$D$. Obviously, in the SU(3) symmetry limit, $\kappa=x=1$,
$\zeta=0$, eq. (\ref{sumrules}) will be $A(D^0\to
K^+\pi^-)=-\tan^2\theta_C A(D^0\to K^-\pi^+)$, and $\delta_{K\pi}$
vanishes. Note that the relation (\ref{sumrules}) does not depend
on the color-suppressed amplitudes ${\cal C}$ and ${\cal
C}^\prime$ because these amplitudes have nothing to do with the
above three decay modes.

In order to go further into the analysis, we need to have
information about the form factors $F_0^{D\to P}(q^2)$. For their
$q^2$ dependence, we adopt the Bauer-Stech-Wirbel model
\cite{BSW}, in which the form factors are assumed to behave as a
monopole, \beq\label{pole}F_0^{D\to P}(q^2)=\frac{F_0^{D\to
P}(0)}{1-q^2/m_*^2},\eeq where $m_*$ is the pole mass with
$m_*=2.47$ GeV for $P=\pi$ and $m_*=2.60$ GeV for $P=K$.
$F_0^{D\to P}(0)$ can be obtained via $F_0^{D\to P}(0)=F_+^{D\to
P}(0)$, since the latter can be measured in semi-leptonic $D^0\to
\pi^- \ell^+ \nu$ and $D^0\to K^-\ell^+\nu$ decays. The new
experimental values from the Belle Collaboration \cite{Belle0604}
give \beqn F_+^{D\to K}(0)=0.695\pm 0.023,~ F_+^{D\to
\pi}(0)=0.624\pm 0.036,\nonumber\eeqn\beqn\label{ratioformfactor}
 F_+^{D\to \pi}(0)/F_+^{D\to
K}(0)=0.898\pm 0.045,\eeqn which are consistent with very recent
results from lattice calculation \cite{lattice} and from the QCD
sum rules calculation \cite{Ball06}. In practice, only the ratio
of these two form factors in (\ref{ratioformfactor}) is needed for
our numerical calculations. By applying it to eq. (\ref{xx}), we
get \beq\label{xx2}x=1.002\pm 0.050, \eeq which is very close to
its value in the SU(3) symmetry limit [the error in (\ref{xx2}) is
due to the uncertainty of $F_+^{D\to \pi}(0)/F_+^{D\to K}(0)$
only]. Although this may be just a numerical coincidence, it seems
that SU(3) symmetry breaking effects are dominated by decay
constants ($f_\pi$ and $f_K$).

\begin{table}[t]\begin{center}\begin{tabular}{ c|c c c c c}
\hline\hline\\ $C_2/C_1$ &$-0.2$&$-0.3$&$-0.4$&$-0.5$&$-0.6$ \\\\
\hline \\$\cos\delta_{K\pi}$ & 0.979$\pm$0.018 &
0.975$\pm 0.021$& 0.969$\pm$0.025 & 0.960$\pm$0.032 & 0.946$\pm0.043$\\
$\delta_{K\pi}$ &
11.7$^\circ\pm 4.9^\circ$& 12.9$^\circ\pm5.3^\circ$ &
14.3$^\circ\pm5.8^\circ$ & 16.2$^\circ\pm6.6^\circ$&18.9$^\circ\pm 7.7^\circ$ \\\hline\\
$\cos\delta_{K\pi}$ &0.983$\pm$0.015 & 0.980$\pm$0.017 &
0.976$\pm$0.021 & 0.970$\pm$0.026 & 0.960$\pm$0.034
\\ $\delta_{K\pi}$& 10.5$^\circ\pm4.6^\circ$ & 11.4$^\circ\pm4.9^\circ$ &
12.6$^\circ\pm5.5^\circ$ & 14.1$^\circ\pm6.1^\circ$&
16.2$^\circ\pm6.9^\circ$
\\\hline\hline\end{tabular}\caption{The relative strong phase
between $D^0\to K^+\pi^-$ and $D^0\to K^-\pi^+$ predicted by eq.
(\ref{deltakpi}) for different values of $C_2/C_1$. ${\cal
B}(D^+\to K^+\pi^0)$ by  BaBar \cite{BABAR06} is used for the
results in the second line; ${\cal B}(D^+\to K^+\pi^0)$ by CLEO
\cite{CLEO0607} is used for the results in the third line. The
sign of $\delta_{K\pi}$ could also be minus.
}\end{center}\end{table}

The numerical predictions of $\delta_{K\pi}$ for different values
of $C_2/C_1$ are displayed in Table 2.  As mentioned above,
$C_2/C_1$ is regarded as a varying parameter. ${\cal B}(D^0\to
K^-\pi^+)$ and ${\cal B}(D^0\to K^+\pi^-)$ are taken from
\cite{PDG06}. Since Particle data group has not given the average
for ${\cal B}(D^+\to K^+\pi^0)$ yet, both of the measurements
listed in eq. (\ref{Babar}) have been used, and the results are
shown in the second and third lines of Table 2, respectively. The
error is due to the uncertainty of $x$ in eq. (\ref{xx2}) and the
uncertainties of experimental branching ratios mentioned above, in
which the contribution from ${\cal B}(D^0\to K^-\pi^+)$ and ${\cal
B}(D^0\to K^+\pi^-)$ is actually very small and can be neglected.
From Table 1, for the relevant scale of $D$ decays, i.e. $\mu$ in
the range of 1.0 $\sim$1.5 GeV, $C_2/C_1$ is about $-0.5\sim
-0.3$. Therefore, a not large but nonzero $\delta_{K\pi}$, whose
magnitude is $10^\circ$ or above, i.e. $\sin\delta\sim \pm 0.2$,
might be expected from the present analysis. The authors of Ref.
\cite{FNP99}, by assuming the existence of nearby resonances for
the $D$ meson, have obtained $\sin\delta_{K\pi}=\pm0.31$, namely,
$\cos\delta_{K\pi}=0.951$ and $\delta_{K\pi}$ is about
$\pm18^\circ$. The other existing hadronic models  which
incorporate SU(3) symmetry breaking effects seem to prefer a small
value of this phase, $\sin\delta\leq 0.2$, with most models giving
$\sin\delta\leq 0.1$ \cite{CC94,BP96} (see Table I in Ref.
\cite{BP96} for details). Unfortunately, the current measurement
of $\delta_{K\pi}$ is very rough \cite{Sun06}, as shown in eq.
(\ref{deltakpiexp}). Employing the asymmetry $R(D)$ measurements
with some theoretical assumptions, another experimental result
$\delta_{K\pi}\approx (3\pm 6\pm 7)^\circ$ with relative small
uncertainty is induced in Ref. \cite{CLEO06}. Both of them are
still consistent with zero.

Finally, we estimate $D\to K\pi$ decay amplitudes from the
currently available data. The three independent complex
phenomenological parameters are chosen as ${\cal T}$, ${\cal C}$,
and ${\cal E}$, not as $a_1^{\rm eff}$, $a_2^{\rm eff}$, and
${\cal E}$, because we will only use the ratio of the form factors
[in eq(\ref{ratioformfactor})] instead of their absolute values in
the analysis. Without loss of generality, ${\cal T}$ is set to be
real. $\delta_C$ ($\delta_E$) is the relative strong phase of
${\cal C}$ (${\cal E}$) to ${\cal T}$. Here we take ${\cal
B}(D^0\to K^-\pi^+)$, ${\cal B}(D^0\to \bar{K}^0\pi^0)$, ${\cal
B}(D^+\to \bar{K}^0\pi^+)$, and ${\cal B}(D^0\to K^+\pi^-)$ given
by Particle data group \cite{PDG06}, together with ${\cal
B}(D^+\to K^+\pi^0)$ by BaBar Collaboration \cite{BABAR06} to
illustrate our numerical calculation. The results of ${\cal T}$,
${\cal C}$, ${\cal E}$, and $|a_2^{\rm eff}/a_1^{\rm eff}|$ are
summarized in Table 3, and other amplitudes ${\cal T}^\prime$,
${\cal C}^\prime$, ${\cal E}^\prime$, and ${\cal A}^\prime$ can be
easily derived using eqs. (\ref{factorization1}),
(\ref{annihilation1}) and (\ref{annihilation3}). Several
observations and remarks are given as follows.

\begin{itemize}

\item The color-suppressed amplitude has a phase $\sim 160^\circ$
relative to the color-allowed amplitude ${\cal T}$, and $|{\cal
C}|$ is effectively enhanced. This means that there could exist
the strongly destructive interference between ${\cal T}$ and
${\cal C}$. We get $a_2^{\rm eff}/a_1^{\rm eff}\simeq 0.56 e^{\pm
i 160^\circ}$, which is insensitive to the value of $C_2/C_1$.
$a_2^{\rm eff}/a_1^{\rm eff}=0.62 e^{-i 152^\circ}$ is obtained in
\cite{cheng03}.

\item The ${\cal E}$ amplitude has a relative phase $\sim
130^\circ$ to ${\cal T}$, $\sim 70^\circ$ to ${\cal C}$.  Its
magnitude is relatively large, and $|{\cal E}|>|{\cal C}|$. This
is contrary to the results in Ref. \cite{cheng03, CR9902}. As
pointed out in \cite{BBNS01}, in general, the weak annihilation
parameter $X_A$ in eq. (\ref{annihilation2}) should be of order
$\ln(m_D/\Lambda$) and $\Lambda$ is a soft scale.  By taking
$\alpha_s\approx 0.5$, $C_1\approx 1.2$, and ${\cal E}=0.325
e^{\pm i 131^\circ}$ for $C_2/C_1=-0.3$, we can roughly estimate
$X_A=4.09 e^{\pm i 122^\circ}$ or $3.66 e^{\pm i 71^\circ}$, which
indicates $|X_A|\sim 2 \ln(m_D/\Lambda)$ with $\Lambda\simeq 0.3$
GeV. $|X_A|=3.84$ has been obtained in \cite{LY05}.

\item Some of our results are not in agreement with the ones in
Refs. \cite{cheng03, CR9902}, since we do not work in the SU(3)
symmetry limit, and we mainly concentrate on $D\to K\pi$ decay
modes in this paper.

\end{itemize}

\begin{table}[t]\begin{center}\begin{tabular}{c c c c c}
\hline\hline\\ $C_2/C_1$~~&~~${\cal T}$ [GeV$^3$]~&~${\cal C}$
[GeV$^3$]~&~${\cal E}$ [GeV$^3$]~&~
$|a_2^{\rm eff}/a_1^{\rm eff}|$\\
\\$-0.3$~~&~~0.417~&~0.289 $e^{\pm i 160^\circ}$~&~0.325 $e^{\mp i 131^\circ}$~&~0.568 \\\\
$-0.4$~~&~~0.445~&~0.306 $e^{\pm i 163^\circ}$~&~0.368 $e^{\mp i
135^\circ}$~&~0.563  \\\\
$-0.5$~~&~~0.485~&~0.334 $e^{\pm i 167^\circ}$~&~0.425 $e^{\mp i
139^\circ}$~&~0.564
\\\hline\hline\end{tabular}\caption{Numerical results of quark-diagram amplitudes
${\cal T}$, ${\cal C}$, ${\cal E}$,  and of $|a_2^{\rm
eff}/a_1^{\rm eff}|$ estimated by using the present data with
different $C_2/C_1$. Only the central values of the magnitude and
the phase are quoted.}\end{center}\end{table}

We return to discuss the asymmetry $R(D^+)$. As pointed out
before, the charged case is not as simple as the neutral case.
Because of \beq\label{rr1}\frac{A(D^+\to K^0 \pi^+)}{A(D^+\to
\bar{K}^0 \pi^+)}=-\tan^2\theta_C \frac{{\cal C}^\prime+{\cal
A}^\prime}{{\cal C}+{\cal T}}=-\tan^2\theta_C\frac{{\cal
C}+C_2/C_1 {\cal E}}{{\cal C}+{\cal T}}, \eeq one cannot simplify
it as a similar analytic relation (\ref{16}) for neutral modes
under ${\cal C}={\cal C}^\prime$ and ${\cal E}={\cal E}^\prime$,
even if including the additional constraint (\ref{annihilation3})
already. However, using the values of amplitudes listed in Table
3, numerically, we  will get
 \beqn\label{rr2} \frac{A(D^+\to K^0
\pi^+)}{A(D^+\to \bar{K}^0
\pi^+)}=\left\{\begin{array}{cc}-\tan^2\theta_C~1.538
e^{\pm i 106^\circ}, & C_2/C_1=-0.3,\\\\
-\tan^2\theta_C~1.532 e^{\pm i 105^\circ}, & C_2/C_1=-0.4,\\\\
-\tan^2\theta_C~1.521 e^{\pm i 103^\circ}, & C_2/C_1=-0.5,
\end{array}\right. \eeqn
which lead to \beqn\label{rdplus}
R(D^+)=\left\{\begin{array}{cc}0.044, & C_2/C_1=-0.3,\\\\0.040,&
C_2/C_1=-0.4,\\\\0.035,& C_2/C_1=-0.5.\end{array} \right.\eeqn
 The present observed value by CLEO Collaboration \cite{CLEO06}
  is $R(D^+)=0.030\pm 0.023\pm 0.025$. Also, the suppression
of $R(D^+)$ comparing with $R(D^0)$ can be understood. From the
definition of $R(D)$ in eq. (\ref{asy0}), one will find it is
proportional to $2 \tan^2\theta_C \cos\delta$, and $\delta$ is the
relative strong phase between the corresponding DCS amplitude and
the CF amplitude. Now it is found that, $\delta$ vanishes in the
$D^0$ case, as shown in eq. (\ref{16}); while $\delta$ is about
$100^\circ$ in the $D^+$ case [see eq. (\ref{rr2})]. Therefore
$R(D^+)$ is suppressed by small $\cos\delta$.

Furthermore, we discuss the possible generalization to the
analysis of SCS $D\to \pi\pi, ~KK$ decays. Consider the ratio
\beq\label{R10}R_1=2\left|\frac{V_{cs}}{V_{cd}}\right|^2
\frac{\Gamma(D^+\to\pi^0\pi^+)}{\Gamma(D^+\to
\bar{K}^0\pi^+)}.\eeq The recent measurement gives $R_1=1.54\pm
0.27$ \cite{BABAR06,PDG06} and $R_1$ should be unity in the SU(3)
symmetry limit. Note that these two modes have only ${\cal T}$ and
${\cal C}$ amplitudes. From our analysis, one can get
\beq\label{R11}
R_1=1.073\left|\frac{F^{D\to\pi}_0(m_\pi^2)}{F^{D\to\pi}_0(m_K^2)}\right|^2\times\left|\frac{1+a_2^{\rm
eff}/a_1^{\rm eff}}{1/x+(f_K/f_\pi) a_2^{\rm eff}/a_1^{\rm
eff}}\right|^2, \eeq where the factor 1.073 is from the
phase-space differences for the $\pi\pi$ and $K\pi$ final states.
Taking $a_2^{\rm eff}/a_1^{\rm eff}\simeq 0.56 e^{\pm i
160^\circ}$ from Table 3, we obtain $R_1\simeq1.44$, in accord
with the recent measurement. Likewise, the ratio \beq\label{R2}
R_2=\frac{\Gamma(D^0\to K^+K^-)}{\Gamma(D^0\to\pi^+\pi^-)}\simeq
1.50 \eeq can also be estimated using Table 3. Unfortunately, this
result is far from the experimental value $\Gamma(D^0\to
K^+K^-)/\Gamma(D^0\to\pi^+\pi^-)=2.82\pm 0.10$ \cite{PDG06},
implying that SU(3) symmetry breaking is still not fully accounted
for. Since now there exist weak-annihilation contributions, in
deriving eq. (\ref{R2}), we have assumed ${\cal
E}_{K^+K^-}=f_K/f_\pi {\cal E}$ and ${\cal E}_{\pi\pi}=f_\pi/f_K
{\cal E}$. This is actually not true because, under this
assumption, the amplitude for the pure weak-annihilation $D^0\to
K^0\bar{K}^0$ decay will vanish, whereas ${\cal B}(D^0\to
K^0\bar{K}^0)=(7.4\pm 1.4)\times 10^{-4}$ experimentally
\cite{PDG06}. Therefore the failure of reproducing the
experimental value in eq. (\ref{R2}) may be unavoidable in the
present framework. In the case of $R_1$, the weak-annihilation
contribution is fortunately absent.  This implies that the above
relations for ${\cal E}_{K^+K^-}$ and ${\cal E}_{\pi\pi}$ need
some corrections, and the weak-annihilation amplitudes should be
carefully investigated when one would like to generalize the
present work to the case of the SCS $D\to \pi\pi, ~K K$ decays
including the pure weak-annihilation mode $D^0\to K^0\bar{K}^0$.

\section{Summary}

We have presented a phenomenological analysis of $D\to K\pi$
decays including both CF and DCS modes. In order to determine all
decay amplitudes for these processes using the present data, a
moderate SU(3) symmetry breaking formalism has been assumed. Our
analysis indicates this assumption works well in $D\to K\pi$
decays. The color-suppressed amplitude is enhanced, and it has a
phase $\sim 160^\circ$ relative to the color-allowed amplitude. A
large weak annihilation amplitude is obtained. Both of the
asymmetries $R(D^0)$ and $R(D^+)$ have been predicted, which are
in good agreement with the experimental data. Our analysis also
shows that a not large but nonzero $\delta_{K\pi}$, which is about
$10^\circ$ or above, might be expected. This means that there is
no good reason to take $\sin\delta_{K\pi}=0$ in the experimental
analysis of $D^0\to K^\pm\pi^\mp$ decays. A precise measurement of
$\delta_{K\pi}$ will be welcome both theoretically and
experimentally.

We would like to point out that, the relation
(\ref{annihilation3}) between ${\cal E}^\prime$ and ${\cal
A}^\prime$ amplitudes, which is important to enable us to
calculate $\delta_{K\pi}$ and estimate the $D\to K\pi$ amplitudes
in this analysis, is a model dependent assumption. Further tests
for this relation will be very useful. But such tests cannot be
performed at present because of the lack of suitable data. On the
other hand, a similar relation between $W$-exchange and
$W$-annihilation amplitudes could occur in SCS $D^0\to K^+K^-$ and
$D^+\to K^+\bar{K}^0$ decays, which might provide an interesting
test. As mentioned above, the present analysis however cannot be
generalized to SCS decays straightforwardly when these transitions
receive contributions from the weak-annihilation amplitudes.
Therefore it would be useful to extend the present framework
assuming (\ref{annihilation3}) to include also SCS decays $D\to
\pi\pi, KK$.
 Since here our main analysis concerns $D\to K\pi$
decays, a further discussion of these issues will be left for
future work.

Very recently, the similar study for $R(D^0)$ has been obtained in
Ref. \cite{JR06}. Since in our framework, we can employ the
currently available data to determine all quark-diagram amplitudes
${\cal T} ({\cal T}^\prime)$, ${\cal C}({\cal C}^\prime)$, ${\cal
E}({\cal E}^\prime)$, and ${\cal A}({\cal A}^\prime)$ including
their relative strong phases,  $R(D^+)$ and $\delta_{K\pi}$ are
also estimated.

\section*{Acknowledgements}
The author is indebted to Gerhard Buchalla  for reading the
manuscript and many stimulating discussions and comments. This
work is supported in part by the Alexander von Humboldt
Foundation.

\end{document}